\documentstyle[11pt]{article}
\begin{document}
\centerline{\bf\large ESTIMATING MASS OF SIGMA-MESON AND STUDY ON}

\vspace{0.6CM}

\centerline{\bf\large APPLICATION OF THE LINEAR SIGMA-MODEL }

\vspace{1cm}

\centerline{Yi-Bing Ding$^1$, Xin Li$^{2}$, Xue-Qian Li$^{2,3}$,
Xiang Liu$^2$, Hong Shen$^2$,} \vspace{0.2cm}

\centerline{ Peng-Nian Shen$^{4}$, Guo-Li Wang$^{2,3}$ and
Xiao-Qiang Zeng$^2$}

\vspace{1cm}

1. Department of Physics, The Graduate School of the Academy of
Sciences of China, Beijing, 10080, China.

2. Department of Physics, Nankai University, Tianjin, 300071,
China.

3. Institute of Theoretical Physics, P.O. Box 2735, Beijing,
100080, China.

4. Institute of High Energy Physics, P.O. Box 918-4, Beijing,
100039, China.

\vspace{1cm}

\begin{center}
\begin{minipage}{12cm}
\noindent {\bf Abstract}

Whether the $\sigma-meson$ ($f_0(600)$) exists as a real particle
is a long-standing problem in both particle physics and nuclear
physics. In this work, we analyze the deuteron binding energy in
the linear $\sigma$ model and by fitting the data, we are able to
determine the range of $m_{\sigma}$ and  also investigate
applicability of the linear $\sigma$ model for the interaction
between hadrons in the energy region of MeV's. Our result shows
that the best fit to the data of the deuteron binding energy and
other experimental data about deuteron advocates a narrow range
for the $\sigma-$meson mass as $520\leq m_{\sigma}\leq 580$ MeV
and the concrete values depend on the input parameters such as the
couplings. Inversely fitting the experimental data, our results
set constraints on the couplings. The other relevant
phenomenological parameters in the model are simultaneously
obtained.
\end{minipage}
\end{center}

\vspace{1cm}

\noindent{I. Introduction}\\

Deuteron is the simplest bound state of nucleons, so that a study
on it may provide us with information of the nuclear force and the
fundamental principle of the strong interaction.

Definitely nucleons are bound together by the strong interaction
whose fundamental theory is QCD. Unfortunately, so far, we are
only confident with the calculation of perturbative QCD at large
momentum transfer i.e. at high energy scales where $\alpha_s$ is
sufficiently small due to the asymptotic freedom. By contrast, for
the lower energy processes, when one evaluates the effects of
hadronization or binding energies, the non-perturbative QCD
effects dominate and the traditional perturbative treatments of
the field theory are no longer applicable. It turns out that one
needs to invoke some phenomenological models in the case. As a
matter of fact, most reasonable models express certain aspects of
the real physical world, but not complete. There is a limited
range for each model, beyond the limit, application of the model
is not legitimate. On other side, for a certain range, all
possible models are equivalent for describing the concerned
physics. Therefore, in some sense,  various models are parallel.
This is understood as we use the Cornell potential, the Richardson
potential or even the logarithmic potential to describe the bound
states of heavy quarks such as $J/\psi$, $\Upsilon$ etc. and
obtain close results with specific parameters. Of course, it by no
means manifests that they are real physics, but just serve as an
effective model and are applicable to the some concerned
processes.

For the quark-bound states, i.e. hadrons, to deal with the
non-perturbative QCD effects, we can use, for example, the QCD sum
rules, potential models, the bag model, as well as the lattice
calculations to evaluate the spectra and other properties. It is
believed that even though $\alpha_s$ is large the quark-gluon
picture is still valid. By contrast, for the nucleon-nucleon
interaction (or hadron-hadron interactions when both of them are
in color singlet), even though it is in principle due to QCD,
there is no single-gluon exchange between the nucleons because
they reside in color-singlet. Instead, just as the familiar Van
der Waals force between moleculas in classical physics, which is
induced by the electric and magnetic multipoles, the strong
interaction between nucleons is the interaction between the
chromo-electric and magnetic multipoles of nucleons. So far, there
is no a successful way to derive an effective form of such strong
interaction at the hadron level from the fundamental QCD theory
yet, but on other side, it is also believed that the chiral
Lagrangian is consistent with the general principles of QCD and
could serve as an effective theory of strong interaction at hadron
level \cite{Georgi}. An alternative version of the chiral
Lagrangian is the linear sigma model where the $\sigma-$particle
stands as an independent resonance. Recently, many theorists and
experimentalists are searching for $\sigma$ which may play an
important role in nuclear physics, because the $\sigma$ meson can
provide reasonable middle-range nuclear force. However, the
present data \cite{PDG,Tor} allow a rather wide range for both the
mass and lifetime of $\sigma-$particle. Thus the existence of
$\sigma$ as a real physical resonance at the hadron spectra of low
energy is a long-standing problem in both particle physics and
nuclear physics. It appears as $f_0(600)$ in the recent data book,
but there still is an acute dispute about its existence. In this
work, we hope to gain more information about this resonance
through studying the deuteron structure.

Deuteron is a familiar object in nuclear physics for many years,
it may be viewed as an ideal lab for probing any possible
mechanism of nuclear force. Many authors have recently studied the
deuteron structure from different angles\cite{stru}, and also
looked for new physics, such as the three-nucleon force via
deuteron breaking up \cite{three}. As indicated above, we trust
that some models are equivalent in principle for the energy range
of a few MeVs, i.e. we can apply any of them. By fitting data, a
set of concerned parameters should be obtained.

Entem and Machleidt \cite{Entem} developed an accurate NN
potential based on the chiral effective theory. Parallel to their
work, in this paper we employ the linear sigma model as an
alternative approach to study the NN potential. The main
difference from \cite{chiral} is that $\sigma$ is an independent
particle here and we can neglect the two-pion exchange
contributions. In realistic physics world $\sigma$ may be a
resonance of certain mass and has the quantum number of two pions
in an S-wave. The calculation now is much simpler because we do
not need to calculate  loop diagrams and deal with the
renormalization. In this model, we derive the NN-effective
potential and  evaluate the spectrum, then compare it with data.
In fact, besides the deuteron binding energy, there are some other
parameters which can be experimentally measured, such as the
mixing fraction of the s- and d-waves, its mean charge radius etc.
Therefore, when we determine the mass of $\sigma$ meson, we need
also to take into account the constraints from the relevant
measurements on the parameters.

In fact, similar to the treatment with the chiral Lagrangian, in
this work all the coefficients at the effective vertices are
derived by fitting data of some physical processes, where all
external particles are supposed to be on their mass-shells.
Therefore, in our derivation of the effective potential, we need
to introduce a reasonable form factor at the effective vertices
which manifests the effects of the inner quark-gluon structure of
nucleons and partly compensate the off-shell effects of the
exchanged mesons at t-channel. Besides we also take into account
the contributions of the light vector mesons to the potential.

This model should be tested by evaluating the deuteron binding
energy which has been measured with high accuracy and we wish to
apply the model along with the obtained parameters to the
discussions on the molecular states \cite{Isgur} which are
important for understanding the hadron spectra, especially at
middle energy ranges.

In this work, we first derive the effective potential between
nucleons. The tensor part of the potential results in a mixing
between the s-wave and d-waves. Namely it not only contributes to
the diagonal elements, but also to the off-diagonal ones of the
hamiltonian matrix. Diagonalizing the matrix, we obtain the mass
eigenvalues, one of them is the deuteron mass and we can argue
that the larger one corresponds to an unstable resonance and does
not exist in the physical world.

This paper is organized as follows, after the introduction, we
derive the formulation and in Sec.III, we present our numerical
results and determine the concerned parameters, finally the last
section is devoted to our conclusion and discussion.\\

\noindent{II. Formulation}\\

Following the traditional method \cite{Lifs} we derive the
effective potential between proton and neutron.

As discussed above we introduce a form factor to compensate the
off-shell effects of the exchanged mesons. At each vertex, the
form factor is written as\cite{lixq}
\begin{equation}
\label{form} {\Lambda^2-M_m^2\over \Lambda^2-q^2},
\end{equation}
where $\Lambda$ is a phenomenological parameter and its value is
near 1 GeV. It is observed that as $q^2\to 0$ it becomes a number
and if $\Lambda\gg M_m$, it turns to be unity, in this case, the
distance is infinitely large and the vertex looks like a perfect
point, so the form factor is simply 1 or a constant. Whereas, as
$q^2\rightarrow \infty$, the form factor approaches to zero, in
this situation, the distance becomes very small, the inner
structure (quark, gluon degrees of freedom) would manifest itself
and the whole picture of hadron interaction is no longer valid, so
the form factor is zero which cuts off the end effects.

To derive an effective potential, one sets $q_0=0$ and writes down
the elastic scattering amplitude in the momentum space and then
carries out a Fourier transformation turning the amplitude into an
effective potential in the configuration space. The best way to
order the operators in the derivation is the Weyl ordering scheme
\cite{Lucha}.

In the linear $\sigma$ model, the effective Lagrangian is

\begin{equation}
L=g\bar{\psi}(\sigma+\gamma_5{\mbox{\boldmath $
\tau$}}\cdot{\mbox{\boldmath $\pi$}})\psi .
\end{equation}

(a) Via single-pion exchange.

The effective vertex is
\begin{eqnarray}
L &=& g\bar{\psi}\gamma_5{\mbox{\boldmath $\tau$}}\cdot {\mbox{\boldmath $\pi$}}\psi \nonumber\\
&=& \sqrt 2g\bigg[\bar p\gamma_5n\pi^++\bar
n\gamma_5p\pi^-+{1\over\sqrt 2}(\bar p\gamma_5p-\bar n\gamma_5
n)\pi^0\bigg],
\end{eqnarray}
where $\psi$ is the wavefunction of nucleon. The scattering
amplitude is
\begin{equation}
M=[-2g^2\bar p\gamma_5 n\bar n\gamma_5 p +g^2\bar p\gamma_5 p\bar
n\gamma_5 n]\cdot {1\over q^2-m_{\pi}^2},
\end{equation}
it is noted that following the traditional treatment, the
wavefunctions of nucleons in the amplitude are that of free
nucleons. Setting $q_0=0$ and including the aforementioned form
factor at the vertices, the potential in the momentum space has a
form
\begin{eqnarray}
V_\pi(\mathbf{q})&=&\frac{\displaystyle g^2}{\displaystyle
4m^2({\mathbf q}^2+m^2_\pi)}
 \bigg[4{\mathbf p}^2_1 -  4({\mbox{\boldmath $\sigma$}}_1\cdot{\mathbf p}_1)({\mbox{\boldmath $\sigma$}}_2\cdot{\mathbf p}_1) +
 2({\mbox{\boldmath $\sigma$}}_2\cdot{\mathbf p}_1)({\mbox{\boldmath $\sigma$}}_1\cdot{\mathbf q}) +
 2({\mbox{\boldmath $\sigma$}}_2\cdot{\mathbf q})({\mbox{\boldmath $\sigma$}}_1\cdot{\mathbf p}_1) \nonumber \\
 &&{}-({\mbox{\boldmath $\sigma$}}_1\cdot{\mathbf q})({\mbox{\boldmath $\sigma$}}_2\cdot{\mathbf q}) -
 2({\mbox{\boldmath $\sigma$}}_2\cdot{\mathbf q})({\mbox{\boldmath $\sigma$}}_2\cdot{\mathbf p}_1) -
 2({\mbox{\boldmath $\sigma$}}_1\cdot{\mathbf p}_1)({\mbox{\boldmath $\sigma$}}_1\cdot{\mathbf q})\bigg]
 \bigg(\frac{\displaystyle \Lambda^2-m^2_\pi}{\displaystyle \Lambda^2+{\mathbf q}^2}\bigg)^2,
\end{eqnarray}
where $m$ is the mass of nucleon. Taking the Fourier
transformation, one has the potential caused by the
single-pion-exchange in the configuration space as
\begin{eqnarray}
\label{pi} V_\pi(r)&=&\frac{\displaystyle g^2}{\displaystyle
4m^2}\bigg[4{\mathbf p}^2_1 f_\pi(r) - 4({\mbox{\boldmath
$\sigma$}}_1\cdot{\mathbf p}_1)({\mbox{\boldmath
$\sigma$}}_2\cdot{\mathbf p}_1)f_\pi(r)
-2i({\mbox{\boldmath $\sigma$}}_2\cdot{\mathbf p}_1)({\mbox{\boldmath $\sigma$}}_1\cdot{\mathbf r})F_\pi(r)\nonumber \\
&&{} - 2iF_\pi(r)({\mbox{\boldmath
$\sigma$}}_2\cdot{\mathbf r})({\mbox{\boldmath
$\sigma$}}_1\cdot{\mathbf p}_1) +({\mbox{\boldmath
$\sigma$}}_1\cdot{\mbox{\boldmath $\nabla$}})({\mbox{\boldmath
$\sigma$}}_2\cdot{\mbox{\boldmath $\nabla$}})f_\pi(r) +
2iF_\pi(r)({\mbox{\boldmath $\sigma$}}_2\cdot{\mathbf r})({\mbox{\boldmath $\sigma$}}_2\cdot{\mathbf p}_1)\nonumber \\
&&{}+ 2i({\mbox{\boldmath
$\sigma$}}_1\cdot{\mathbf p}_1)({\mbox{\boldmath
$\sigma$}}_1\cdot{\mathbf r})F_\pi(r)\bigg].
\end{eqnarray}
This potential is not hermitian, we need to re-order the operator
${\bf p}$ and the functions of $r$ and the standard treatment is
the Weyl ordering procedure \cite{Lucha}. The resultant effective
potential is
\begin{eqnarray}
\lefteqn{V_{\pi}(r)_{Weyl}= }\nonumber \\
& & \frac{g^{2}}{4 m^{2}} \bigg\{ [f_{\pi}(r) {\mathbf p}_{1}^{2} +
{\mathbf p}_{1} f_{\pi}(r) {\mathbf p}_{1} + {\mathbf p}_{1}^{2}
f_{\pi}(r)] - [f_{\pi}(r)({\mbox{\boldmath $\sigma$}}_{1} \cdot
  {\mathbf p}_{1})({\mbox{\boldmath $\sigma$}}_{2} \cdot {\mathbf p}_{1}) \nonumber \\
  & & {} + ({\mbox{\boldmath $\sigma$}}_{1} \cdot
  {\mathbf p}_{1})f_{\pi}(r)({\mbox{\boldmath $\sigma$}}_{2} \cdot {\mathbf p}_{1}) +({\mbox{\boldmath $\sigma$}}_{1} \cdot
  {\mathbf p}_{1})({\mbox{\boldmath $\sigma$}}_{2} \cdot {\mathbf p}_{1}) f_{\pi}(r)]- i [({\mbox{\boldmath $\sigma$}}_{2} \cdot {\mathbf p}_{1})({\mbox{\boldmath $\sigma$}}_{1}
  \cdot {\mathbf r}) F_{\pi}(r) \nonumber\\
  & & {}+ F_{\pi}(r)({\mbox{\boldmath $\sigma$}}_{1}
  \cdot {\mathbf r})({\mbox{\boldmath $\sigma$}}_{2} \cdot {\mathbf p}_{1})]-i[({\mbox{\boldmath $\sigma$}}_{1} \cdot {\mathbf p}_{1})({\mbox{\boldmath $\sigma$}}_{2}
  \cdot {\mathbf r}) F_{\pi}(r) \nonumber+ F_{\pi}(r)({\mbox{\boldmath $\sigma$}}_{2}
  \cdot {\mathbf r})({\mbox{\boldmath $\sigma$}}_{1} \cdot {\mathbf p}_{1})] \nonumber \\
  & &{} + i[({\mbox{\boldmath $\sigma$}}_{1} \cdot {\mathbf p}_{1})({\mbox{\boldmath $\sigma$}}_{1}
  \cdot {\mathbf r}) F_{\pi}(r) \nonumber+ F_{\pi}(r)({\mbox{\boldmath $\sigma$}}_{1}
  \cdot {\mathbf r})({\mbox{\boldmath $\sigma$}}_{1} \cdot {\mathbf p}_{1})]+i[({\mbox{\boldmath $\sigma$}}_{2} \cdot {\mathbf p}_{1})({\mbox{\boldmath $\sigma$}}_{2}
  \cdot {\mathbf r}) F_{\pi}(r) \nonumber \nonumber \\
  & & {}+ F_{\pi}(r)({\mbox{\boldmath $\sigma$}}_{2}
  \cdot {\mathbf r})({\mbox{\boldmath $\sigma$}}_{2} \cdot {\mathbf p}_{1})]+ [({\mbox{\boldmath $\sigma$}}_{1} \cdot
  {\mbox{\boldmath $\nabla$}})({\mbox{\boldmath $\sigma$}}_{2} \cdot {\mbox{\boldmath $\nabla$}})
  f_{\pi}(r)]\bigg\},
\end{eqnarray}
where
\begin{eqnarray}
  f_\pi(r)& = &\frac{\displaystyle e^{-m_\pi r}}{\displaystyle 4\pi r} -
 \frac{\displaystyle e^{-\Lambda r}}{\displaystyle 4\pi r} +
 \frac{\displaystyle (m_\pi^2-\Lambda^2)e^{-\Lambda r}
 }{\displaystyle 8\pi\Lambda}\nonumber\\
 F_\pi(r) & = &\frac{1}{\displaystyle r}\frac{\displaystyle \partial}{\displaystyle \partial
  r}f_\pi(r).
  \end{eqnarray}
And it is the form we are going to use in the later part of the
work.\\

(b) Via $\sigma$ and $\rho$ and $\omega$ exchanges.

The effective vertices are respectively
\begin{equation}
L_{\sigma}=g\bar\psi \psi\sigma,
\end{equation}
\begin{equation}
L_{\rho}=g_{_{NN}\rho}\bar\psi\gamma_{\mu}\tau^a\psi A^{a\mu}\;\;
a=1,2,3,
\end{equation}
and for $\omega-$vector meson it is
\begin{equation}
L_{\omega}=g_{_{NN}\omega}\bar\psi\gamma_{\mu}\psi \omega^{\mu}.
\end{equation}

Neglecting some technical details, one can easily derive the
corresponding effective potentials as
\begin{eqnarray}
V_{\sigma}(r) &=& g^{2} \bigg\{
-f_{\sigma}(r)+\frac{[f_{\sigma}(r) {\mathbf p}_{1}^{2} +
{\mathbf p}_{1} f_{\sigma}(r) {\mathbf p}_{1} + {\mathbf p}_{1}^{2}
f_{\sigma}(r)]}{4
m^{2}}\nonumber\\
& -& \frac{[{\mbox{\boldmath$\nabla$}}^{2} f_{\sigma}(r)]}{4
m^{2}} {}+ \frac{i [{\mathbf p}_{1} \cdot {\mathbf r} F_{\sigma}(r)
+ F_{\sigma}(r) {\mathbf r} \cdot {\mathbf p}_{1}] }{2 m^{2}} -
\frac{({\mathbf L}\cdot{\mathbf S}) F_{\sigma}(r)}{2 m^{2}} \bigg\},
\end{eqnarray}
where
\begin{eqnarray}
  f_\sigma(r)&=&\frac{\displaystyle e^{-m_\sigma r}}{\displaystyle 4\pi
r}-\frac{\displaystyle e^{-\Lambda r}}{\displaystyle 4\pi
r}+\frac{\displaystyle (m_\sigma^2-\Lambda^2)e^{-\Lambda r}
 }{\displaystyle 8\pi\Lambda}\nonumber\\
  F_\sigma(r)&=&\frac{1}{\displaystyle r}\frac{\displaystyle
\partial}{\displaystyle \partial r}f_\sigma(r).
   \end{eqnarray}
and
\begin{eqnarray}
\label{rho}
\lefteqn{V_{\rho}(r) =} \nonumber\\
& & {} \frac{g_{_{NN}\rho}^{2}}{4 m^{2}}  \bigg \{ 4 m^{2}
f_{\rho}(r) +[{\mbox{\boldmath$\nabla$}}^{2}f_{\rho}(r)]
-\frac{[f_{\rho}(r) {\mathbf p}_{1}^{2} + {\mathbf p}_{1}
f_{\rho}(r) {\mathbf p}_{1} + {\mathbf p}_{1}^{2} f_{\rho}(r)]}{2}
+2 F_{\rho}(r){\mathbf L}\cdot {\mathbf S}\nonumber\\
& & {} - i [{\mathbf p}_{1} \cdot {\mathbf r} F_{\rho}(r)+
F_{\rho}(r) {\mathbf r} \cdot {\mathbf p}_{1} ] +\frac{5}{2}
[f_{\rho}(r)({\mbox{\boldmath $\sigma$}}_{1} \cdot
  {\mathbf p}_{1})({\mbox{\boldmath $\sigma$}}_{2} \cdot {\mathbf p}_{1}) \nonumber  + ({\mbox{\boldmath $\sigma$}}_{1}
  \cdot
  {\mathbf p}_{1})f_{\rho}(r)({\mbox{\boldmath $\sigma$}}_{2} \cdot {\mathbf p}_{1}) \nonumber\\
  & & {}+({\mbox{\boldmath $\sigma$}}_{1} \cdot
  {\mathbf p}_{1})({\mbox{\boldmath $\sigma$}}_{2} \cdot {\mathbf p}_{1}) f_{\rho}(r)]
  +\frac{5i}{2}[F_{\rho}(r)({\mbox{\boldmath $\sigma$}}_{1} \cdot {\mathbf r})({\mbox{\boldmath $\sigma$}}_{2}
  \cdot {\mathbf p}_{1})+({\mbox{\boldmath $\sigma$}}_{2} \cdot {\mathbf p}_{1})
({\mbox{\boldmath $\sigma$}}_{1} \cdot {\mathbf r}) F_{\rho}(r)] \nonumber\\
& & {}
 +\frac{5i}{2}[F_{\rho}(r)({\mbox{\boldmath $\sigma$}}_{2} \cdot {\mathbf r})({\mbox{\boldmath $\sigma$}}_{1}
 \cdot {\mathbf p}_{1})+({\mbox{\boldmath $\sigma$}}_{1} \cdot {\mathbf p}_{1})
  ({\mbox{\boldmath $\sigma$}}_{2} \cdot {\mathbf r})F_{\rho}(r)]
 - [({\mbox{\boldmath $\sigma$}}_{1}
 \cdot {\mbox{\boldmath$\nabla$}})({\mbox{\boldmath $\sigma$}}_{2}
 \cdot {\mbox{\boldmath$\nabla$}})f_{\rho}(r)]\nonumber\\
 & & {}+\frac{1}{2}[({\mbox{\boldmath $\sigma$}}_{1} \times {\mbox{\boldmath $\sigma$}}_{2})
 \cdot {\mathbf p}_{1}({\mbox{\boldmath $\sigma$}}_{1} \cdot {\mathbf r})F_{\rho}(r)
 +F_{\rho}(r)({\mbox{\boldmath $\sigma$}}_{1} \cdot {\mathbf r}) ({\mbox{\boldmath $\sigma$}}_{1}
 \times {\mbox{\boldmath $\sigma$}}_{2}) \cdot {\mathbf p}_{1}] \nonumber\\
 & & {}+\frac{1}{2}[F_{\rho}(r)({\mbox{\boldmath $\sigma$}}_{1} \times {\mbox{\boldmath $\sigma$}}_{2})
 \cdot {\mathbf r}({\mbox{\boldmath $\sigma$}}_{2} \cdot {\mathbf p}_{1})
 +({\mbox{\boldmath $\sigma$}}_{2} \cdot {\mathbf p}_{1}) ({\mbox{\boldmath $\sigma$}}_{1}
 \times {\mbox{\boldmath $\sigma$}}_{2}) \cdot {\mathbf r} F_{\rho}(r)]\nonumber\\
 & & {}-\frac{1}{2}[F_{\rho}(r)({\mbox{\boldmath $\sigma$}}_{1} \times {\mbox{\boldmath $\sigma$}}_{2})
 \cdot {\mathbf r}({\mbox{\boldmath $\sigma$}}_{1} \cdot {\mathbf p}_{1})
 +({\mbox{\boldmath $\sigma$}}_{1} \cdot {\mathbf p}_{1}) ({\mbox{\boldmath $\sigma$}}_{1}
 \times {\mbox{\boldmath $\sigma$}}_{2}) \cdot {\mathbf r} F_{\rho}(r)]\nonumber\\
 & & {}-\frac{1}{2}[({\mbox{\boldmath $\sigma$}}_{1} \times {\mbox{\boldmath $\sigma$}}_{2})
 \cdot {\mathbf p}_{1}({\mbox{\boldmath $\sigma$}}_{2} \cdot {\mathbf r})F_{\rho}(r)
 +F_{\rho}(r)({\mbox{\boldmath $\sigma$}}_{2} \cdot {\mathbf r}) ({\mbox{\boldmath $\sigma$}}_{1}
 \times {\mbox{\boldmath $\sigma$}}_{2}) \cdot {\mathbf p}_{1}]
\bigg\},
\end{eqnarray}
where
\begin{eqnarray}
 f_\rho(r)&=&\frac{\displaystyle e^{-m_\rho r}}{\displaystyle 4\pi
r}-\frac{\displaystyle e^{-\Lambda r}}{\displaystyle 4\pi
r}+\frac{\displaystyle (m_\rho^2-\Lambda^2)e^{-\Lambda r}\nonumber
 }{\displaystyle 8\pi\Lambda}\\
   F_\rho(r)&=&\frac{1}{\displaystyle r}\frac{\displaystyle \partial}{\displaystyle \partial
   r}f_\rho(r).
    \end{eqnarray}

The potential induced by exchanging $\omega$ vector meson is in
close analogy to $V_{\rho}$, with only $m_{\rho}$ being replaced
by $m_{\omega}$. We neglect the  contributions by exchanging
heavier mesons.

The total effective potential is the sum of these potentials
induced respectively via exchanging  $\pi,\;\sigma$ and $\rho,\;
\omega$ mesons.
$$V_{eff}(r)=V_{\pi}(r)+V_{\sigma}(r)+V_{\rho}(r)+V_{\omega}(r).$$
After  simple manipulations, we can write the $V_{eff} (r)$ as a
sum containing a central potential term  $V_C(r)$, a
spin-dependent central potential part $V_S(r)\; {\mbox{\boldmath
$\sigma$}}_1\cdot{\mbox{\boldmath $\sigma$}}_2 $, a spin-orbit
coupling part $V_{LS}(r)\; {\bf L}\cdot {\bf S} $, and a tensor
interaction part $V_{T}(r)\; S_{12} $. The tensor operator $
S_{12}=[3({\mbox{\boldmath $\sigma$}}_1\cdot{\mathbf \hat{r}})
            ({\mbox{\boldmath $\sigma$}}_2\cdot{\mathbf \hat{r}})
           -{\mbox{\boldmath $\sigma$}}_1\cdot{\mbox{\boldmath $\sigma$}}_2] $
is known to admix the s-wave and d-wave states,
so we separate out the tensor part of the potential as
\begin{eqnarray}
  V_T(r) S_{12}
 &=& \bigg[
\bigg(- \frac{\displaystyle e^{-\Lambda r}\Lambda^3
g_{_{NN}\rho}^2}{\displaystyle 64m^2\pi}+\frac{\displaystyle
e^{-\Lambda r}\Lambda m_\rho^2g_{_{NN}\rho}^2}{\displaystyle
64m^2\pi}-\frac{\displaystyle 3e^{-\Lambda
r}g_{_{NN}\rho}^2}{32m^2\pi r^3}+\frac{\displaystyle 3e^{-m_\rho
r}g_{_{NN}\rho}^2}{32m^2\pi r^3}-\frac{\displaystyle 3e^{-\Lambda
r}\Lambda g_{_{NN}\rho}^2}{32m^2\pi r^2} \nonumber \\ &&
 {}+\frac{\displaystyle 3e^{-m_\rho r}m_\rho
g_{_{NN}\rho}^2}{32m^2\pi r^2}-\frac{\displaystyle 3e^{-\Lambda
r}\Lambda^2 g_{_{NN}\rho}^2}{64m^2\pi r}+\frac{\displaystyle
e^{-\Lambda r}m_\rho^2 g_{_{NN}\rho}^2}{64m^2\pi
r}+\frac{\displaystyle e^{-m_\rho r}m_\rho^2
g_{_{NN}\rho}^2}{32m^2\pi r}\bigg)\nonumber \\
&&+ \bigg(-\frac{\displaystyle e^{-\Lambda
r}\Lambda^3g_{_{NN}\omega}^2}{\displaystyle 64m^2\pi}
 {}+\frac{\displaystyle e^{-\Lambda r}\Lambda
m_\omega^2g_{_{NN}\omega}^{2}}{\displaystyle
64m^2\pi}-\frac{\displaystyle 3e^{-\Lambda
r}g_{_{NN}\omega}^2}{32m^2\pi r^3}+\frac{\displaystyle
3e^{-m_\omega r}g_{_{NN}\omega}^2}{32m^2\pi
r^3}\nonumber \\
&&-\frac{\displaystyle 3e^{-\Lambda r}\Lambda
g_{_{NN}\omega}^2}{32m^2\pi r^2}+\frac{\displaystyle 3e^{-m_\omega
r}m_\omega g_{_{NN}\omega}^2}{32m^2\pi r^2}  -\frac{\displaystyle
3e^{-\Lambda r}\Lambda^2 g_{_{NN}\omega}^2}{64m^2\pi
r}+\frac{\displaystyle e^{-\Lambda r}m_\omega^2
g_{_{NN}\omega}^2}{64m^2\pi r}\nonumber \\
&&+\frac{\displaystyle e^{-m_\omega r}m_\omega^2
g_{_{NN}\omega}^2}{32m^2\pi r}\bigg) \bigg] S_{12},
\end{eqnarray}
and treat it as a perturbation later.

Now, the Schr\"odinger equations for s-wave state $(^3 \rm{S}_1)$
and d-wave state $(^3 \rm{D}_1)$ have the following forms
\begin{equation}
\label{schs}
 \bigg[ \frac{{\bf p}^2}{2\mu}+V_{C}(r)+V_{S}(r)\bigg]\Psi_s({\mathbf{r}})
 =E_s\Psi_s({\mathbf{r}}),
\end{equation}
\begin{equation}
\label{schd}
 \bigg[ \frac{{\bf p}^2}{2\mu}+V_{C}(r)+V_{S}(r)-3V_{LS}(r)-2V_T(r)\bigg]\Psi_d({\mathbf{r}})
 =E_d\Psi_d({\mathbf{r}}),
\end{equation}
where $\mu$ is the reduced mass, here it is ${1\over 2}M_p={1\over
2}M_n$. Solving these equations, we obtain the eigen-energies and
eigen-wavefunctions of the s-wave and d-wave states. Then the
tensor part, which results in a mixing fraction between the s- and
d-waves, would be taken in as a perturbation.

We have a modified hamiltonian as
\begin{equation}
\label{full} H_{full}=\left( \begin{array}{ll}
E_s      & \Delta \\
\Delta^* & E_d
\end{array} \right).
\end{equation}
Here $E_s$ and $E_d$ are the eigenvalues of s- and d-waves
obtained by solving the Schr\"odinger equations (\ref{schs}) and  (\ref{schd}),
$\Delta$ is defined as
$$\Delta=<s|V_T(r)S_{12}|d>.$$
Diagonalizing $H_{full}$, we  obtain the real eigenvalues and
identify the lower one to be the binding energy of deuteron (we
will discuss this issue in the last section again), at the same
time we can determine the fraction of the d-wave in the
eigenfunction
$$|\Psi>=c_1|s>+c_2|d>.$$

\noindent{III. The numerical results}

\vspace{0.2cm}

The input parameters include the masses $m_{\pi}=0.139$ GeV,
$m_{\rho}=0.77$ GeV, $m_{\omega}=0.783$ GeV, $m_p\approx
m_n=0.938$ GeV, and the couplings
$g_{_{NN}\sigma}=g_{_{NN}\pi}=13.5$, $g_{_{NN}\rho}=3.25$
\cite{a}. In the exact SU(3) limit, by a simple quark counting,
$g_{_{NN}\omega}=3 g_{_{NN}\rho}$, but our later numerical results
show that this relation does not lead to a satisfactory solution.
For a comparison, we will take three different $g_{_{NN}\omega}$
values which are $g_{_{NN}\rho}$, $2g_{_{NN}\rho}$ and
$3g_{_{NN}\rho}$ respectively and present the results in tables
1,2 and 3.

Even though we do not have direct information about
$g_{NN\sigma}$, $g_{NN\sigma}=g_{NN\pi}$ is imposed by the linear
sigma model. $\Lambda$ is a free parameter in the form factor, it
is generally believed that its value is in a range around 1 GeV.
Therefore, one can vary it within a reasonable region around 1
GeV. Here we take it as a free parameter close to 1 GeV and adjust
it to fit the data of deuteron. When we carry out the numerical
computations, by imposing the experimental data as constraints
which include the binding energy of deuteron, a mixing between s-
and d-waves and the charge radius etc., it is found that a unique
$\Lambda-$value for both s- and d-waves cannot lead to a
reasonable solution set which can meet the data. Therefore we take
separate values for s- and d-waves, more discussions will be given
in the last section.

The main task of this work is to find the mass of
$\sigma-$particle, it stands as free a parameter in the present
calculation and we will scan the possible region and finally
determine its value by fitting data. As well known, $\sigma-$meson
is a rather wide resonance, it is so far not well experimentally
determined yet, the data allow a large range as the mass being
$400\sim 1200$ MeV and width being $600\sim 1000$ MeV\cite{PDG}.
For deriving the effective potential the width is irrelevant,
because the t-channel is a deep space-like region, so the width is
not involved, thus in this work we cannot determine the width from
the static properties of deuteron. If one wishes to study the
width, he must calculate the production rate or decay (lifetime)
of deuteron which would be more difficult and we will deal with it
in our later work.

The experimental value of the binding energy of deuteron is
\cite{Bonn}
$$E_b^{exp}=-2.224575\; {\rm MeV}.$$

Our strategy is to fit the binding energy which is very accurately
measured and  serves as a rigorous criterion for our computation.
Then we achieve a reasonable mixing fraction between the s- and
d-waves which has also been experimentally confirmed. As
aforementioned, to fulfil all the criteria, a unique $\Lambda-$
value is not enough, thus we set $\Lambda_s$ and $\Lambda_d$ for
s- and d-waves respectively when we solve the Schr\"odinger
equations. We present our numerical results in Table 1.

\vspace{1cm}

\vspace{0.2cm}

\begin{center}
\begin{tabular}{|c||c|c|c|c|c|c|c|} \hline
$m_{\sigma}$ & $\Lambda_{s}$ & $\Lambda_{d}$ & $E_{S}$ & $E_{D}$ & $\Delta$ &  & $\bar{r}$ \\
($\times 10^{-1}$GeV)&(GeV)&(GeV)&(MeV)&(MeV)&(MeV)&\raisebox{2.3ex}[0pt]{$(\frac{\displaystyle c_{d}}{\displaystyle c_{s}})^2$}&(fm)\\
\hline \hline
4.70&0.59&1.35&-1.97&-0.29&0.71&0.13&3.54\\
\hline
4.80&0.60&1.40&-2.00&-0.36&0.64&0.12&3.47\\
\hline
4.90&0.61&1.46&-2.04&-0.40&0.58&0.10&3.39\\
\hline
5.00&0.62&1.52&-2.08&-0.43&0.52&0.08&3.31\\
\hline
5.10&0.63&1.58&-2.11&-0.44&0.46&0.07&3.23\\
\hline
5.20&0.64&1.65&-2.13&-0.48&0.40&0.05&3.16\\
\hline
5.30&0.65&1.71&-2.15&-0.51&0.35&0.04&3.08\\
\hline
5.40&0.66&1.78&-2.17&-0.54&0.29&0.03&3.01\\
\hline
5.50&0.67&1.84&-2.19&-0.55&0.25&0.02&2.94\\
\hline
5.60&0.69&1.91&-2.22&-0.58&0.20&0.01&2.88\\
\hline
\end{tabular}
\end{center}

\vspace{0.2cm}

Table 1.
$g_{NN\pi}=g_{NN\sigma}=13.5$,$g_{NN\rho}=3.25$,$g_{NN\omega}=g_{NN\rho}$
All the other parameters are obtained  by fitting the binding
energy $E_b^{exp}=-2.22$ GeV

\vspace{0.2cm}

With another set of couplings, we have\\

\begin{center}
\begin{tabular}{|c||c|c|c|c|c|c|c|} \hline
$m_{\sigma}$ & $\Lambda_{s}$ & $\Lambda_{d}$ & $E_{S}$ & $E_{D}$ & $\Delta$ &  & $\bar{r}$ \\
($\times 10^{-1}$GeV)&(GeV)&(GeV)&(MeV)&(MeV)&(MeV)&\raisebox{2.3ex}[0pt]{$(\frac{\displaystyle c_{d}}{\displaystyle c_{s}})^2$}&(fm)\\
\hline \hline
5.10&0.64&1.75&-1.70&-0.30&1.00&0.27&3.54\\
\hline
5.20&0.65&1.84&-1.80&-0.33&0.90&0.22&3.45\\
\hline
5.30&0.66&1.92&-1.89&-0.36&0.79&0.18&3.36\\
\hline
5.40&0.67&2.02&-1.97&-0.40&0.68&0.14&3.28\\
\hline
5.50&0.68&2.11&-2.04&-0.43&0.58&0.10&3.19\\
\hline
5.60&0.69&2.21&-2.09&-0.46&0.48&0.08&3.10\\
\hline
5.70&0.70&2.30&-2.13&-0.50&0.39&0.05&3.02\\
\hline
5.80&0.71&2.40&-2.17&-0.51&0.31&0.03&2.92\\
\hline
\end{tabular}
\end{center}
\vspace{0.2cm}

Table 2.
$g_{NN\pi}=g_{NN\sigma}=13.5$,$g_{NN\rho}=3.25$,$g_{NN\omega}=2g_{NN\rho}$
All the other parameters are obtained  by fitting the binding
energy $E_b^{exp}=-2.22$ GeV

\vspace{0.2cm}

Then for a comparison we give another set of results which is
listed in table 3.\\

\begin{center}
\begin{tabular}{|c||c|c|c|c|c|c|c|} \hline
$m_{\sigma}$ & $\Lambda_{s}$ & $\Lambda_{d}$ & $E_{S}$ & $E_{D}$ & $\Delta$ &  & $\bar{r}$ \\
($\times 10^{-1}$GeV)&(GeV)&(GeV)&(MeV)&(MeV)&(MeV)&\raisebox{2.3ex}[0pt]{$(\frac{\displaystyle c_{d}}{\displaystyle c_{s}})^2$}&(fm)\\
\hline \hline
3.50&0.53&0.79&-2.16&-0.44&3.81&0.64&4.45\\
\hline
4.00&0.57&1.03&-2.08&-0.40&3.46&0.62&3.97\\
\hline
4.50&0.61&1.46&-1.92&-0.25&2.72&0.55&3.67\\
\hline
5.00&0.65&2.14&-1.84&-0.19&1.86&0.43&3.44\\
\hline
5.50&0.69&3.07&-1.68&-0.09&1.02&0.24&3.36\\
\hline
\end{tabular}
\end{center}

Table 3.
$g_{NN\pi}=g_{NN\sigma}=13.5$,$g_{NN\rho}=3.25$,$g_{NN\omega}=3g_{NN\rho}$
All the other parameters are obtained  by fitting the binding
energy $E_b^{exp}=-2.22$ GeV.

\vspace{0.2cm}

Obviously, even though all the three sets of couplings can lead to
results which coincide with the data on the binding energy, when
we consider other experimental constraints, especially the s-wave
and d-wave mixing fraction of about 4\%, the set given in table 3
is not favored.

It is noted that to fit the binding energy of deuteron and
consider the constraint of the s- and d-wave mixing, the estimated
mass of $\sigma$ meson deviates for different parameter sets by
about 50 MeV. We will further discuss this issue in the last
section. And we have also noted that for each parameter set,
beyond the range of $m_{\sigma}$, we cannot find any reasonable
solution for the binding energy and the mixing
parameter $(C_d/C_s)^2$.\\

\noindent{IV. Discussions and conclusion}

\vspace{0.2cm}

Nowadays, the hadron   structure is still a hot topic in particle
physics,  we wonder if the observed hadron spectra from light to
heavy can be fully interpreted in the quark model. Among all the
light hadrons, the scalar $\sigma\; (0^+,\; f_0(600))$ may take
the most noticeable position. Dispute about its existence as a
physical resonance has been on for a long time. A recent
observation of the narrow resonance ($2.32\;GeV/c^2$) by the BABAR
collaboration  \cite{BABAR} has triggered a new tide to discuss
the whole family of the $0^+$ particles\cite{Beveren} where
$\sigma$ is the leading one. One of the most striking questions is
whether they can be attributed to the building of $q\bar q$ system
or some of them may be molecular states. Indeed, the measured
resonance peaks $f_0(980)$, $a_0(980)$ are interpreted as $K\bar
K$ molecula \cite{Isgur}, Rujula, Georgi and Glashow suggest that
$\Psi(4.04)$ is a $D^*\bar D^*$ molecula \cite{Ruj} etc. To
further investigate such molecular states, one needs to know the
interaction between the meson-constituents in the molecular states
and derive the effective potential. In that case, the $\sigma$
particle may be the main agent to mediate the interaction as well
as pion and other light  vector mesons. Thus to test the
properties of $\sigma$ becomes the most important task in the
whole study.

There are many ways to study $\sigma$, among them the structure of
deuteron  provides us with direct information about the $\sigma$
particle. Deuteron is the simplest bound state of nucleons, in
general, it is a "molecular" state of two fermions (proton and
neutron). Because its spectrum is accurately measured, it is an
ideal probe for the applicable mechanism and theoretical
framework.

In this work, we employ the linear $\sigma$ model where the
$\sigma$ is a real physical particle whereas in the chiral
Lagrangian approach its role is replaced by the two-pion-exchange
picture. At the effective vertices, we retain a form factor with a
free parameter $\Lambda$ which is determined by fitting data.

When we consider the constraint from the experimental observation
that a certain fraction of about 4\%  d-wave mixes with the
s-wave, we fit the binding energy and the results are listed in
Table 1. As aforementioned, a unique $\Lambda$ value cannot lead
to a solution for the binding energy and mixing of around 4\%.
Thus we take two parameters $\Lambda_s$ and $\Lambda_d$
corresponding to s- and d-waves respectively. In fact, $\Lambda$
represents the inner structure of the constituents of deuteron and
the effective interaction between them. For s- and d-waves, the
orbital angular momenta are different, and the momentum
distributions of the constituents in the s- and d-states behave
differently, namely the effective binding could deviate from each
other. The different $\Lambda-$ values for s- and d-waves would
manifest the difference in effective interaction for different
angular momenta states.

The effective coupling $g_{_{NN}\sigma}=g_{_{NN}\pi}=13.5$ is
taken in our numerical computations, but other possibilities are
also noted that in literature, a different value of the coupling
is also commonly used as
$g_{_{NN}\sigma}=g_{_{NN}\pi}=10.1$\cite{Shen}. To investigate the
influence of different couplings, we repeat all the numerical
computations with the coupling being replace by
$g_{_{NN}\sigma}=g_{_{NN}\pi}=10.1$, the results are not in a good
shape, namely no reasonable solutions for both the binding energy
and mixing fraction can be reached. Thus this model does not
advocate smaller coupling $g_{_{NN}\sigma}=g_{_{NN}\pi}=10.1$.

It is worth noticing that the results somewhat depend on the
coupling $g_{_{NN}\omega}$. Indeed, both the $\rho$ and $\omega$
mesons provide an effective short-range expelling interaction
between nucleons, by the SU(3) symmetry, $g_{_{NN}\phi}=0$ and
$g_{_{NN}\omega}\approx 3g_{_{NN}\rho}$ \cite{a}, if a simple
quark counting rule is used. But the relation does not need to be
taken seriously, because the SU(3) symmetry is obviously broken.
In fact, in the literature, quite different relations between
$g_{_{NN}\omega}$ and $g_{_{NN}\rho}$ have been adopted. One would
rather use the loose relation as $g_{_{NN}\omega}=\alpha
g_{_{NN}\rho}$ where $\alpha$ is treated as a free parameter and
is to be determined by fitting data. In the text, we adopt three
different $\alpha$ values to be 1, 2 and 3 respectively and
present corresponding results in the tables. We observe that all
the results corresponding to the three relations can lead to
solutions which perfectly coincide with the data of the deuteron
binding energy, but while considering the s-wave and d-wave mixing
fraction $|c_d/c_s|^2$ as a constraint, one can immediately notice
that for larger $\alpha$ values the fraction becomes unbearably
large. It indicates that smaller $\alpha$ values are more favored,
and it may also manifest the SU(3) breaking in this case or in
other words, for an effective theory for the NN interaction at
very low energy (near zero), $g_{_{NN}\omega}\approx
g_{_{NN}\rho}$.

As a conclusion of this work, the data on deuteron advocates the
mass range of the $\sigma$ particle around 520 to 580 MeV which is
consistent with the present data and the value commonly used in
the phenomenological nuclear physics. Meanwhile we obtain a
fraction of the d-wave in the total wave function as about 4$\sim$
10\% and it also coincides with the observation. One question
might be raised that if deuteron is a mixed state of the s- and
d-waves with the lower eigen-energy, where is the  bound state of
the higher energy. Since deuteron is the unique stable nucleus
consisting of a proton and a neutron, no other bound state has
been experimentally observed, it implies that the bound state of
higher eigen-energy is unstable and easy to dissociate, so does
not exist in the nature. Indeed, our calculation shows that as the
binding energy of deuteron is $-2.22$ MeV, another eigen-energy is
very close to zero.

It has been reported that the newly measured value of the mass and
width of $\sigma-$meson as
$$m_{\sigma}=390^{+60}_{-36}\;{\rm MeV}\;\;\;{\rm and}\;\;\;
\Gamma_{\sigma}=282^{+77}_{-50}\;{\rm MeV}\cite{Wu}.$$ Huo et al.
recently also discussed on the $\sigma$ properties in $J/\psi$
decays \cite{Huo}. Their value of the mass of $\sigma-$meson might
be a bit smaller than the generally expected. Instead, in this
work we use the deuteron data to investigate the mass. Because of
the measurement errors, we take several $m_{\sigma}$ values and
recalculate $E_b$.


In general, our result indicates that the linear $\sigma$ model
may apply to determine phenomenological quantities of deuteron and
inversely by studying these quantities, one can achieve further
information about this long-expected particle $f_0(600)$.   As
many theorists and experimentalists concern its existence and
search for it at various experimental facilities and processes, we
believe that its existence will be confirmed soon and its
properties including mass and lifetime will be fixed with much
higher accuracy in near future. Then we will be able to further
study our mechanism and the parameter $\Lambda$. These results can
be compared with other data  of wider energy ranges for getting a
better insight into the strong interactions. That achievements
would enrich our knowledge about the interaction mediated by the
$\sigma$ meson  and then we will apply it to other areas of high
energy physics and nuclear physics. However, from the other side,
we find that the results depend on the adopted parameters which
are obtained by fitting the data of nucleon scattering and given
in different works of literature. Further study on $\sigma$ meson
would provide more definite information about its mass and then we
can continue to investigate the linear $\sigma$ model.

\vspace{0.5cm}

\noindent Acknowledgements:

This work is partially supported by the National Natural Science
Foundation of China. We would like to thank Prof. P.Z. Ning for
helpful discussions and introduction to the study of deuteron in
nuclear physics. We also benefit from discussions with Prof. F.
Wang and Prof. X.F. L\"u.

\end{document}